\documentclass{SLT2024}
\pdfoutput=1

\interspeechcameraready 

\title{LANGUAGE BIAS IN SELF-SUPERVISED LEARNING FOR AUTOMATIC SPEECH RECOGNITION}

\name[affiliation={1,2}]{Edward}{Storey}
\name[affiliation={1}]{Naomi}{Harte}
\name[affiliation={2}]{Peter}{Bell}

\address{
  $^1$Sigmedia Lab, School of Engineering, Trinity College Dublin, Ireland\\
  $^2$Centre for Speech Technology Research, School of Informatics, The University of Edinburgh, UK
  }

\email{storeyed@tcd.ie, nharte@tcd.ie, peter.bell@ed.ac.uk}


\begin{document}

\maketitle
\begin{abstract}
Self-supervised learning (SSL) is used in deep learning to train on large datasets without the need for expensive labelling of the data. Recently, large Automatic Speech Recognition (ASR) models such as XLS-R have utilised SSL to train on over one hundred different languages simultaneously. However, deeper investigation shows that the bulk of the training data for XLS-R comes from a small number of languages. Biases learned through SSL have been shown to exist in multiple domains, but language bias in multilingual SSL ASR has not been thoroughly examined. In this paper, we utilise the Lottery Ticket Hypothesis (LTH) to identify language-specific subnetworks within XLS-R and test the performance of these subnetworks on a variety of different languages. We are able to show that when fine-tuning, XLS-R bypasses traditional linguistic knowledge and builds only on weights learned from the languages with the largest data contribution to the pretraining data.
\end{abstract}

\section{INTRODUCTION}
Pretrained Automatic Speech Recognition (ASR) models are now able to process numerous languages within a single model with low error rates even on low-data languages \cite{babu2021xlsr, whisper}. It has been consistently shown that fine-tuning large pretrained models provides the best accuracy on a single language task, when compared to smaller models \cite{Huang_CrossLingual, Toshniwal_MultiLingLSTM}. To achieve this accuracy, multilingual ASR models contain hundreds of millions of parameters and must be pretrained on hundreds of thousands of hours of speech data. Self-supervised learning (SSL) is increasingly utilised to process such large amounts of data without the need for expensive labelling and has become common in open-source ASR \cite{wav2vec, hsu2021hubert, chen2022wavlm}. However, previous commercial and open-sourced supervised learning models have been shown to produce higher errors when presented with speech data outside the domain of their training data \cite{koenecke2020racial, towards_inclusive}. It is therefore imperative that the data and training strategies employed in SSL ASR are diligently explored.

Recent studies have shown that the data that SSL ASR models are trained on can impact downstream training. Boito \textit{et al}. \cite{sslbias_gender} have experimented with wav2vec 2.0 by pretraining on data that was not balanced equally in gender and then fine-tuning to gender-balanced data. They found that if pretraining data was not gender-balanced, the error would increase when fine-tuning to gender-balanced speech. Meng \textit{et al}. \cite{sslbias_speechrate} showed that biasing pretraining data towards slow speech rates for SSL transformer-based models improves downstream accuracy, whereas fast speech in pretraining has a performance drop. To show the impact of this pretraining Fuckner \textit{et al}. \cite{sslbias_Dutch} investigate wav2vec 2.0 performance when fine-tuned to Dutch speech data. They find that Whisper \cite{whisper}, a supervised learning model trained on multilingual data, outperforms wav2vec 2.0 \cite{wav2vec} which is pretrained only using English data. Zhang \textit{et al}. \cite{zhang2023fast} pretrain the wav2vec 2.0 architecture on a single language. They analysed data from multiple languages and selectively included the utterances that contained the most language similarity with the target language into the pretraining data. With this approach, they could attain baseline results with significantly reduced data and training time.  These studies show that biasing the pretraining data towards certain data domains can affect the downstream performance of SSL ASR models.

With this in mind it is vital to analyse the data commercial open-source SSL ASR models are trained on. SSL excels when trained on hundreds of thousands of hours of data and the most abundant source of speech data is English language data. Several models \cite{wav2vec, hsu2021hubert} used for multilingual ASR are trained solely on English data such as Librispeech \cite{librispeech} or Libri-Light \cite{librilight}. Other SSL models are pretrained on multilingual data \cite{babu2021xlsr, XLSR-53}, but they will often contain more English data than other languages. Multilingual Librispeech (MLS) \cite{MLS} is commonly used as a source of multilingual data. However, out of the 50k hours of data in MLS, 44k are English speech data. 

This study aims to identify the impact that the imbalance of English data in SSL pretraining has on open-source commercial ASR as, to our knowledge, this has not been deeply explored before. In order to achieve this, we require techniques that identify low-level behaviours of large deep learning models. Model compression techniques such as model distillation \cite{distillation}, low-rank adaptation \cite{hu2021lora} and model pruning \cite{han2015learning} all look to exploit specific behaviours in large pretrained deep learning models in order to reduce the size of a model without affecting performance. They can therefore be useful tools to analyse low-level behaviours of our large SSL ASR models.

Through the use of the Lottery Ticket Hypothesis (LTH) \cite{frankle2018LTH} to prune deep learning models, several studies have shown the existence of language-specific weight groupings or ``subnetworks'' within large ASR models \cite{PARP}. These subnetworks of weights contribute more to learning on a downstream language than other groups. These language-specific subnetworks have been used to improve accuracy in sparsely pruned networks for data of the same or related languages \cite{Lu_Language_adap_xlsr, yang2023learning}. Each subnetwork's performance on unrelated languages has not been extensively studied prior to the work presented here. For this study we chose to evaluate XLS-R \cite{babu2021xlsr}, XLS-R is an open-source SSL ASR model based on the same architecture as wav2vec 2.0 \cite{wav2vec}. We evaluate the 300 million parameter version that was pretrained through SSL on multilingual data from 128 different languages. This paper will utilise LTH in order to identify language-specific weights and subnetworks within XLS-R and evaluate the extent to which language balance in the pretraining data affects bias towards or against performance on various languages in downstream fine-tuning tasks.

\section{BACKGROUND}
\subsection{Self-Supervised Learning and XLS-R} \label{ssec:SSL_and_XLSR}
Among highly cited open source SSL ASR models, wav2vec 2.0 and HuBERT were trained on Librispeech \cite{librispeech} and Libri-Light \cite{librilight}, which only include English language data \cite{wav2vec, hsu2021hubert}. XLS-R \cite{babu2021xlsr}, however, is a large SSL ASR model pretrained on 128 different languages. It is built on the same architecture as wav2vec 2.0 but expanded to 24 transformer layers in the encoder. 

Like other multilingual SSL ASR models, such as its predecessor XLSR-53 \cite{XLSR-53}, the pretraining data contains more English data than any other language, due to the largest datasets included in the XLS-R pretraining data \cite{MLS, wang-etal-2021-voxpopuli}. English language speech data accounts for approximately 15.9\% of the total training data. Additionally, the first 24 languages account for 98\% of the total data XLS-R was trained on \cite{babu2021xlsr}. We chose XLS-R for this study due to its variety of languages and also the imbalance in the number of hours for each language in the pretraining data. The number of hours and percentage total data for each language included in this study are outlined in Table \ref{tab:XLSR_Data_Splits}, a breakdown of the number of hours per language can be found in the original XLS-R paper \cite{babu2021xlsr}.
\begin{table}[h]
\centering
\begin{tabular}{lll}
\cline{2-3} \\[-2ex]
 & \multicolumn{2}{c}{Proportion of The Pretraining Data} \\ \cline{1-3} 
\multicolumn{1}{r|}{Language} & \multicolumn{1}{c}{No of Hours} & \multicolumn{1}{c}{Percentage (\%)} \\ \hline
\multicolumn{1}{r|}{English} & \multicolumn{1}{c}{69.5k} & \multicolumn{1}{c}{15.9} \\ 
\multicolumn{1}{r|}{German} & \multicolumn{1}{c}{25.4k} & \multicolumn{1}{c}{5.8} \\ 
\multicolumn{1}{r|}{French} & \multicolumn{1}{c}{24k} & \multicolumn{1}{c}{5.5} \\ 
\multicolumn{1}{r|}{Spanish} & \multicolumn{1}{c}{22.3k} & \multicolumn{1}{c}{5.1} \\ 
\multicolumn{1}{r|}{Polish} & \multicolumn{1}{c}{20.9k} & \multicolumn{1}{c}{4.8} \\ 
\multicolumn{1}{r|}{Catalan} & \multicolumn{1}{c}{691} & \multicolumn{1}{c}{0.16} \\ \hline
\multicolumn{1}{r|}{Total Hours} & \multicolumn{1}{c}{162.8k} & \multicolumn{1}{c}{37.33}   
\end{tabular}
\caption{\textbf{The proportions to which each language tested in this study exists within the XLS-R pretraining data} XLS-R pretraining data is 436k hours in total\\[-6.5ex]}
\label{tab:XLSR_Data_Splits}
\end{table}
\subsection{Model Compression and Network Pruning}
Model compression is an area of study that aims to reduce the size of large pretrained models while preserving their accuracy. All of these techniques take advantage of low-level changes in weights over the course of training. So, for our purposes, we can adapt them as tools to monitor how weights behave across the model when subjected to a variety of data. There are multiple active areas of research when it comes to compressing large deep-learning models such as model distillation \cite{distillation}, low-rank adaptation \cite{hu2021lora} and model pruning \cite{frankle2018LTH}. 

Model pruning allows for the removal or zeroing of certain weights across the model if they are not deemed necessary for downstream tasks \cite{han2015learning}. The Lottery Ticket Hypothesis (LTH) \cite{frankle2018LTH} states that unstructured pruning of weights within a pretrained network can uncover ``winning tickets''. These are subnetworks of weights that contribute more to learning downstream tasks than other groups. For our purposes, analysing these subnetworks can give insight into which weights are most active when different language data stimuli are introduced to a model. 

\subsection{Language-Specific Subnetworks}
Several studies have shown that language-specific subnetworks can be obtained when pruning large multilingual ASR models. Lu \textit{et al}. \cite{Lu_Language_adap_xlsr} test joint training with multiple language-specific subnetworks. They find this approach has less degradation on the initial high-resource language performance than when pruning with a language-agnostic approach. Similarly, in 2023, Yang \textit{et al}. \cite{yang2023learning} show that language-specific subnetworks have better performance than language-agnostic approaches.

In 2021, Lai \textit{et al}. \cite{PARP} studied language-specific networks in wav2vec 2.0 \cite{wav2vec} and XLSR-53 \cite{XLSR-53}. They use the Intersection Over Union (IOU) to show the overlap of weights in different language-specific subnetworks, finding high overlap between different subnetworks, other than those that are randomly generated. They also test the performance of language-specific subnetworks on various different languages. They find that there is a large variability in performance depending on which language-specific subnetwork is used for which downstream language. However, this study does not explore an English subnetwork and concentrates on the language-agnostic benefits of their own pruning algorithm. In this paper, we explore how pretraining data affects language-specific subnetworks and then assess their performance on a wide array of languages. Approaching language-specific subnetworks with this method will give us insight into how different pretraining languages impact multilingual SSL ASR performance.

\section{METHOD}
\subsection{Model} 
XLS-R \cite{babu2021xlsr} is an ASR model pretrained through self-supervised learning on 128 separate languages and has high performance on many languages when fine-tuned. The smallest iteration of XLS-R with 300M parameters was selected for this paper. 

\begin{figure}[b]
  \centering
  \includegraphics[scale=0.2]{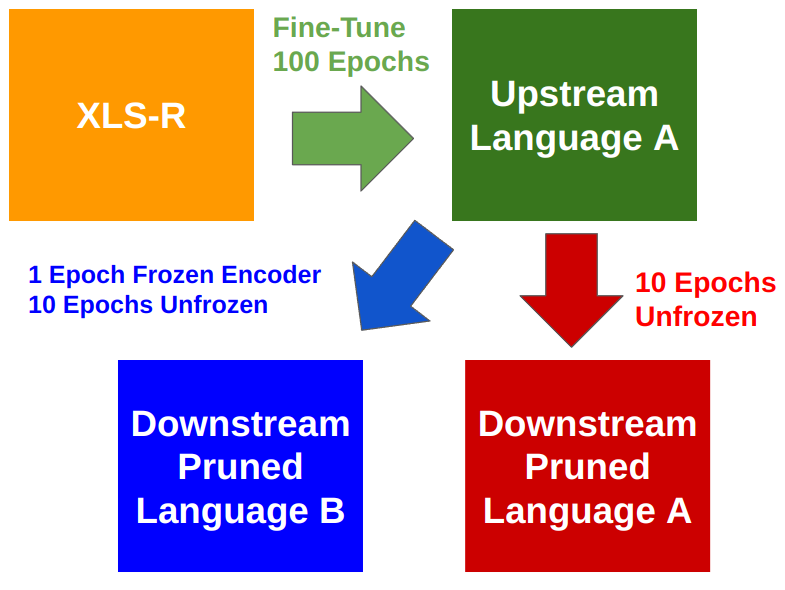}
  \caption{\textbf{Training pipeline for all models} XLS-R is fine-tuned to an upstream language. We then prune and train the downstream task for 10 epochs. If the upstream and downstream languages do not match we freeze the encoder and train for 1 extra epoch before unfreezing the encoder and training for 10 epochs}
  \label{fig:Training_Strategy}
\end{figure}
\subsection{Data}
The FLEURS: FEW-Shot Learning Evaluation of Universal Representations of Speech dataset \cite{FLEURS} contains data from 101 languages. FLEURS contains approximately 12 hours of data per language. For our upstream languages, we selected English, German, French, Spanish and Polish. These languages are all Indo-European languages from three sub-groups: Germanic, Latin and Slavic \cite{bauer2007linguistics}. They are all high-data languages within the XLS-R pretraining data, as seen in Table \ref{tab:XLSR_Data_Splits}. All of these languages were tested downstream alongside Asturian and Xhosa. Asturian and Xhosa are languages unseen by XLS-R in its pretraining \cite{babu2021xlsr}. Asturian is a language spoken in Northern Spain and like Spanish is a Latin language \cite{Asturian_Muñiz-Cachón_2018}. Xhosa is a language spoken in South Africa with no linguistic relationship to the other languages in these experiments \cite{bauer2007linguistics}. Finally, Catalan is a low-data language in XLS-R pretraining data and is a Latin language from Spain.

\subsection{Pruning}
In all of these experiments, we apply L1-norm unstructured one-shot global weight pruning to the encoder of XLS-R. Global pruning was used across the transformer-based encoder to target more valuable connections that may exist across layers and not force the pruning of a single layer more than is necessary \cite{frankle2018LTH}. 

\subsection{Training Strategies}
To obtain subnetworks for each language tested, we must first train XLS-R on our upstream language. We use the same hyperparameters for XLS-R as Rouditchenko \textit{et al}. \cite{FLEURS_XLSR_WHISPER} but we differ by training on 8 2080ti GPUs. First, we train XLS-R for 100 epochs and select the point in training with the lowest CTC loss on the validation set for the final upstream model. Then to create our downstream model we prune the upstream model to the target sparsity and train it on either the same or a new language.

When the downstream language is the same as the upstream language (i.e. English further trained on English), we train the downstream model for 10 epochs at sparsities increasing in steps of 10\%. For downstream languages that do not match the upstream language (i.e. an English model trained on downstream Spanish), we freeze the encoder for one epoch of training before unfreezing and continuing to train for the 10 epochs. We evaluate the final models' performance based on Character Error Rate (CER) as in \cite{FLEURS_XLSR_WHISPER}. All training pathways can be seen in Figure \ref{fig:Training_Strategy}.

\def\figscale{0.83}
\def\capsscale{1}
\begin{figure}[t]
  \centering
  \captionsetup{width=\capsscale\linewidth}
  \includegraphics[width=\linewidth]{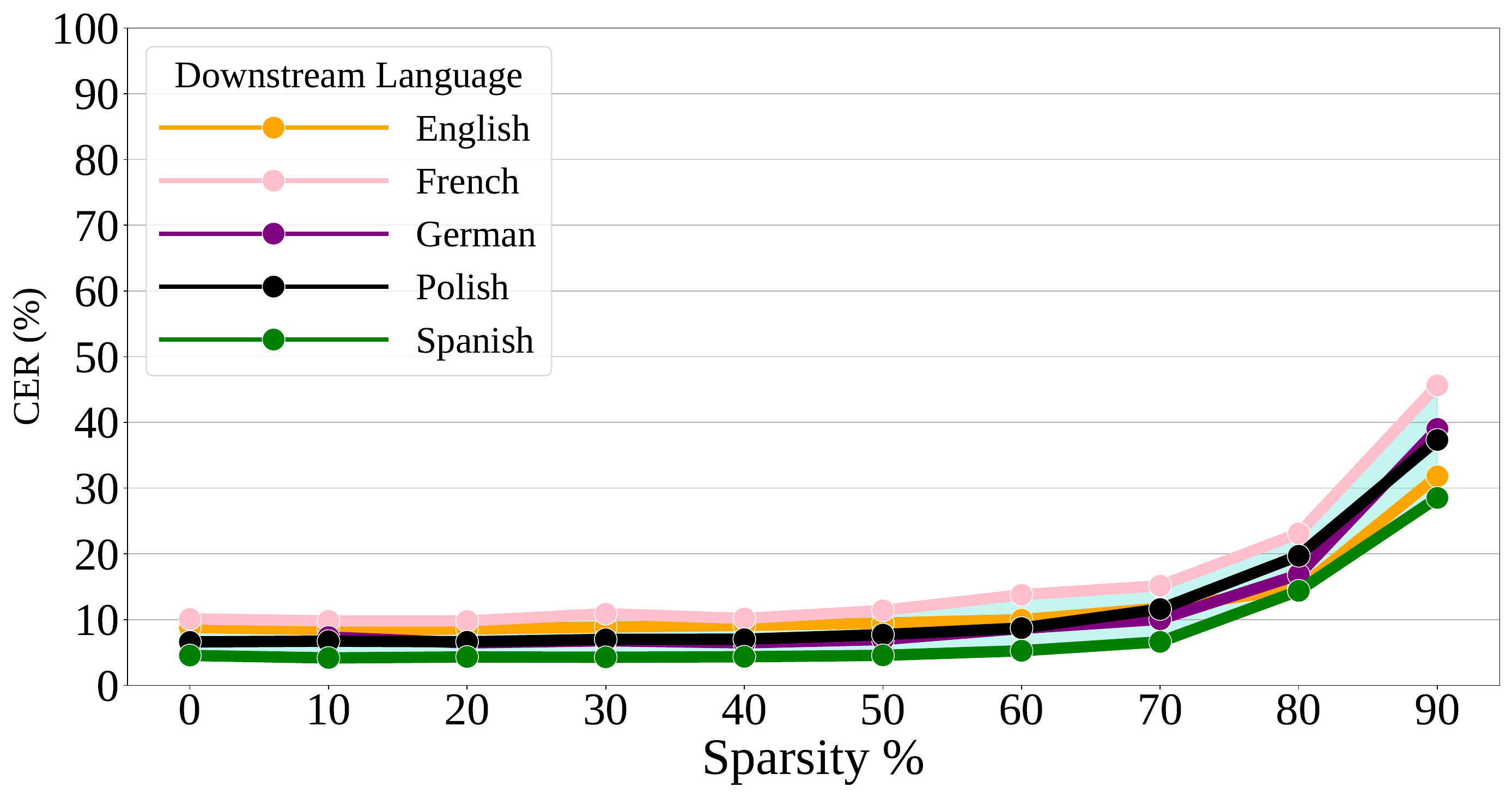}
  \caption{\textbf{Upstream English to multiple downstream Languages} an English upstream model is fine-tuned to downstream English, French, German, Polish and Spanish while pruning from 0\% up to 90\% sparsity}
  \label{fig:English_to_0to90}
\end{figure}
\begin{figure}[t]
  \includegraphics[width=\linewidth]{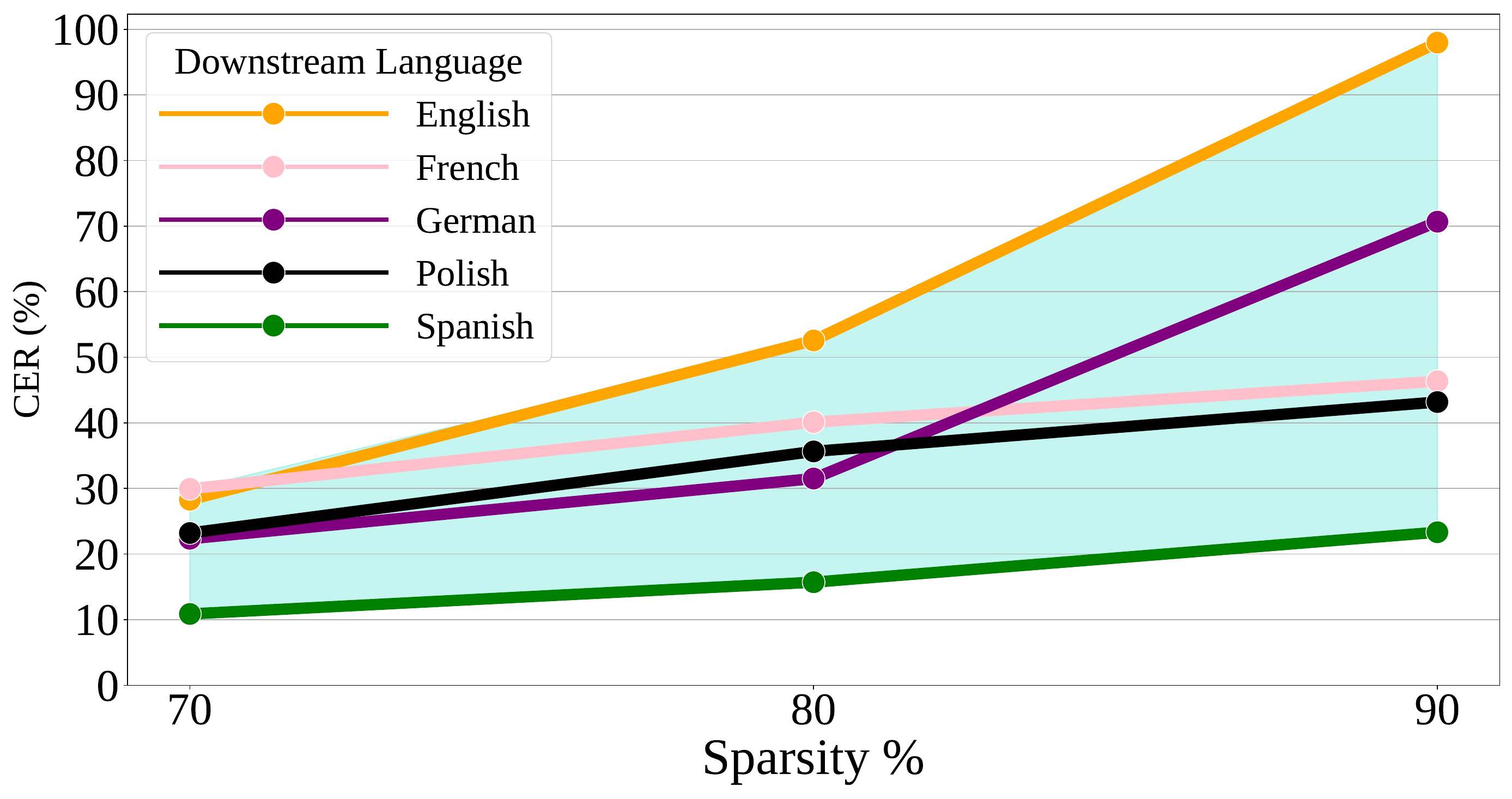}
  \caption{\textbf{Upstream Spanish to five downstream languages} the Spanish upstream model is fine-tuned to downstream English, French, German, Polish and Spanish at 70\%, 80\% and 90\% sparsities}
  \label{fig:Upstream_Spanish}
\end{figure}
\begin{figure}[t]
  \includegraphics[width=\linewidth]{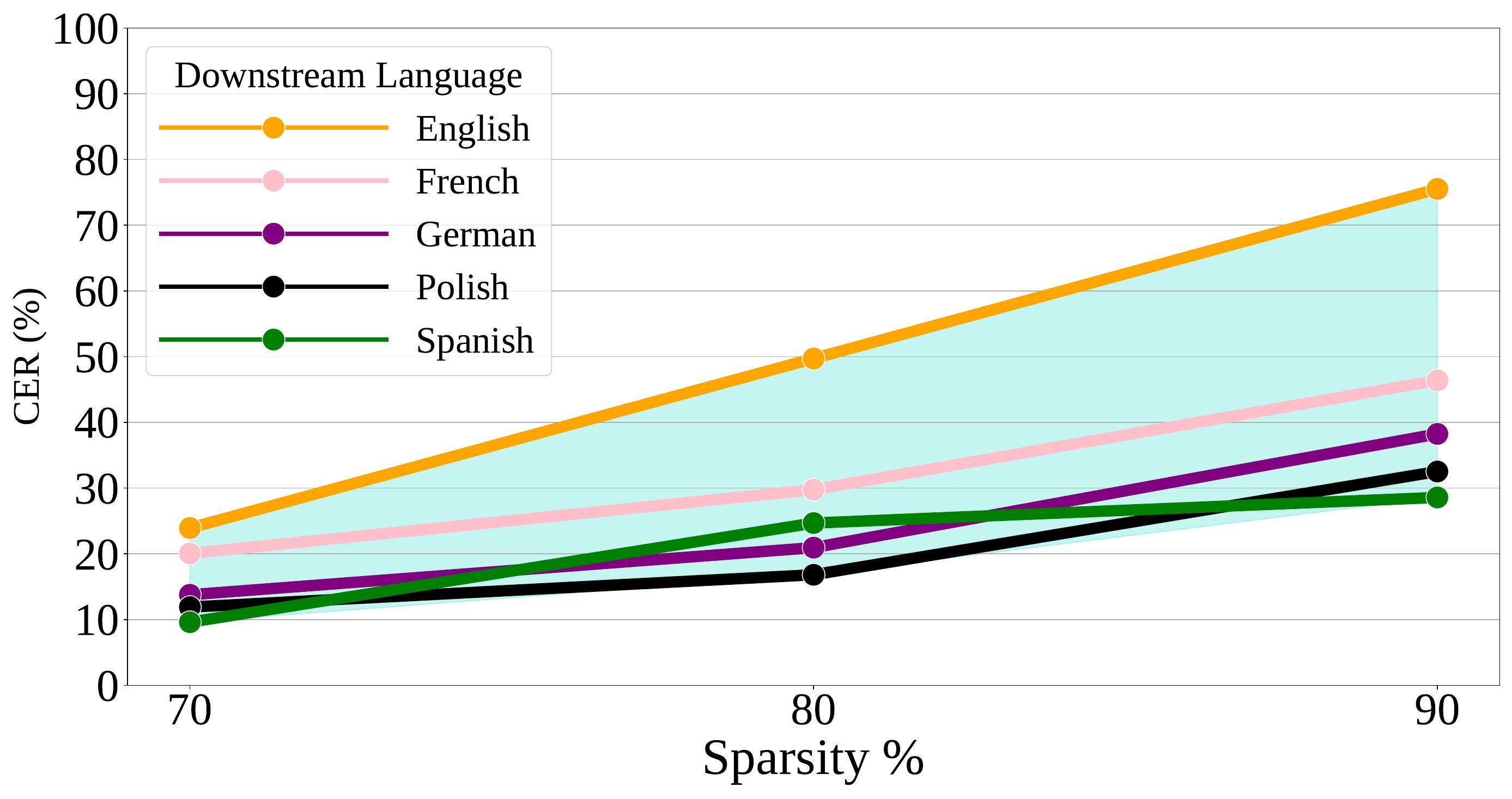}
  \caption{\textbf{Upstream Polish to five downstream Languages} the Polish upstream model is fine-tuned to downstream English, French, German, Polish and Spanish at 70\%, 80\% and 90\% sparsities}
  \label{fig:Upstream_Polish}
\end{figure}

\section{RESULTS} \label{sec:Results}

\subsection{Language Specific Subnetworks}
Figure \ref{fig:English_to_0to90} shows XLS-R trained on English and pruned to sparsities starting at 0\% up to 90\% sparsity in increments of 10\%. Figure \ref{fig:English_to_0to90} corroborates findings in previous papers \cite{ding2022audio, Losses_NEURIPS2022_83d349b6, Unstructured_Speech} that show that sparse networks can achieve minimal degradation to the baseline unpruned accuracy of a model up to 50\% or 60\%. The impact of language-specific subnetworks is more pronounced in sparsities at or above 70\%. As such our findings throughout the rest of this paper will concentrate on 70\%, 80\% and 90\% sparsities.

Figure \ref{fig:English_to_0to90}, Figure \ref{fig:Upstream_Spanish} and Figure \ref{fig:Upstream_Polish} show the same experiments performed with different upstream languages. Figure \ref{fig:English_to_0to90} shows upstream English as a base for the language-specific subnetworks, Figure \ref{fig:Upstream_Spanish} shows upstream Spanish and Figure \ref{fig:Upstream_Polish} shows upstream Polish, three languages from separate Indo-European language families \cite{bauer2007linguistics} and all high-data within the XLS-R pretraining data. Across the three figures, we can observe a clear trend of English as an outlier to the other languages-specific subnetworks. Figure \ref{fig:English_to_0to90} has the lowest CER across all languages, even when the downstream language matches the upstream language as with Spanish in Figure \ref{fig:Upstream_Spanish} and Polish in Figure \ref{fig:Upstream_Polish}. Figure \ref{fig:Upstream_Spanish} and Figure \ref{fig:Upstream_Polish} both show substantial increases in CER on English at 80\% and 90\% sparsity when compared to the other language-specific subnetworks tested. This may suggest that the English language subnetwork is more effective for downstream training. Adverse to the findings in \cite{PARP} Figure \ref{fig:Upstream_Spanish} and Figure \ref{fig:Upstream_Polish} suggest that not training on an English subnetwork is in fact detrimental to downstream fine-tuning, even on languages unrelated to English.

\begin{figure}[t]
  \centering
  \includegraphics[width=\linewidth]{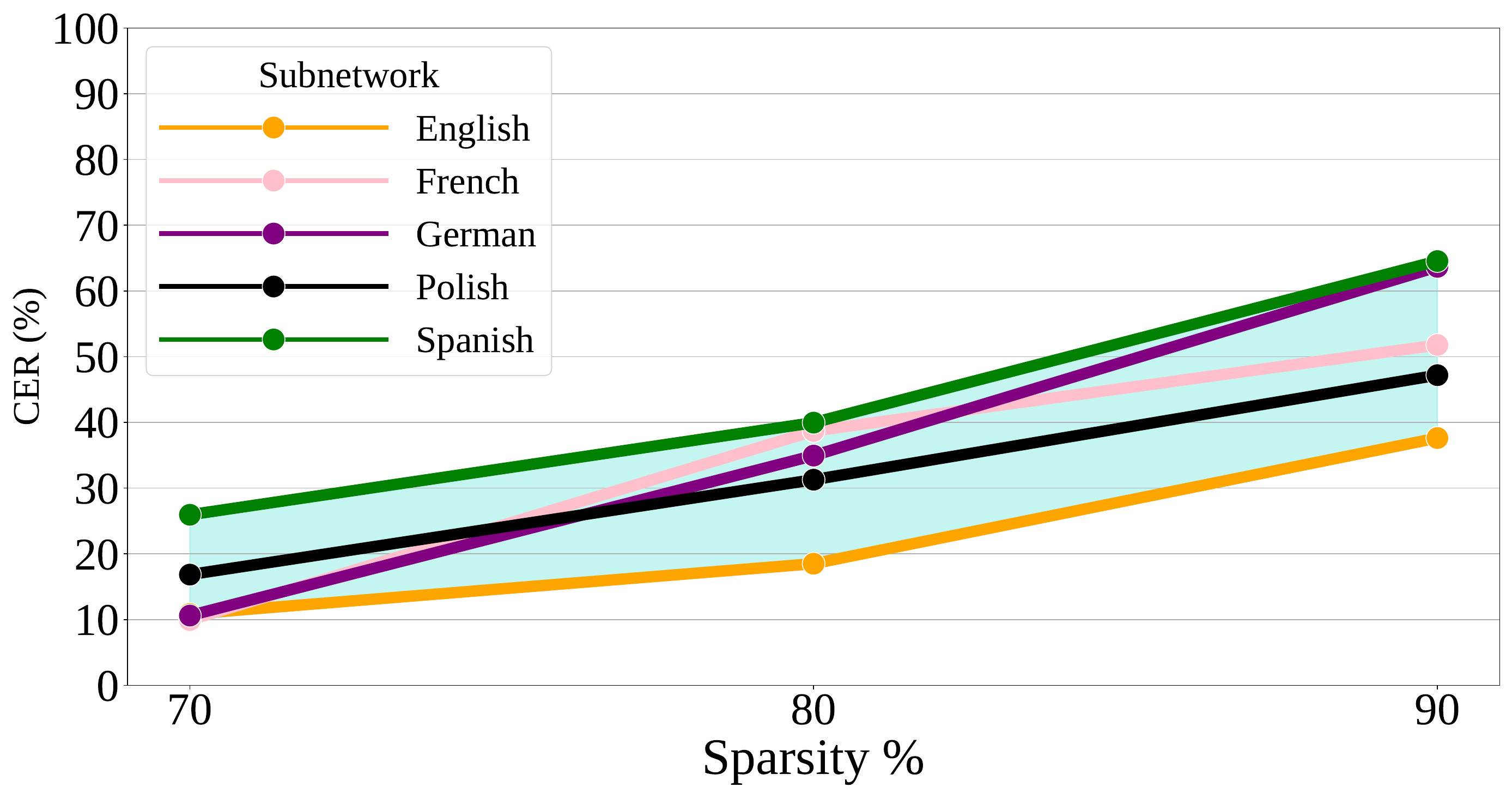}
  \caption{\textbf{Mean average results for each language-specific subnetwork when tested on all other downstream languages} each subnetwork is fine-tuned to the four other languages, these results are then averaged and plotted at 70\%, 80\% and 90\% sparsity\\[-3ex]}
  \label{fig:Language_Averages}
\end{figure}

We further corroborate these findings in Figure \ref{fig:Language_Averages} which shows the average CER for each upstream model. To gauge how well these models generalise, we exclude the language the model was trained on for the results in Figure \ref{fig:Language_Averages}, i.e. the average performance for the English language-specific subnetworks fine-tuned to French, German, Polish and Spanish at each sparsity. We see in Figure \ref{fig:Language_Averages} that the CER at 80\% and 90\% sparsity is the lowest among all subnetworks with an absolute difference of 26.92\% CER at 90\% sparsity between the highest and lowest error and 21.46\% CER at 80\%. While the English CER at 70\% is not the lowest among subnetworks, it is within 1\% error from the lowest of the French subnetwork and 15.09\% less than the highest CER at 70\% sparsity for the Spanish subnetwork. This shows that across all other languages, English language-specific subnetworks perform as well or better than the other subnetworks.

\begin{figure}[h]
  \centering
  \includegraphics[width=\linewidth]{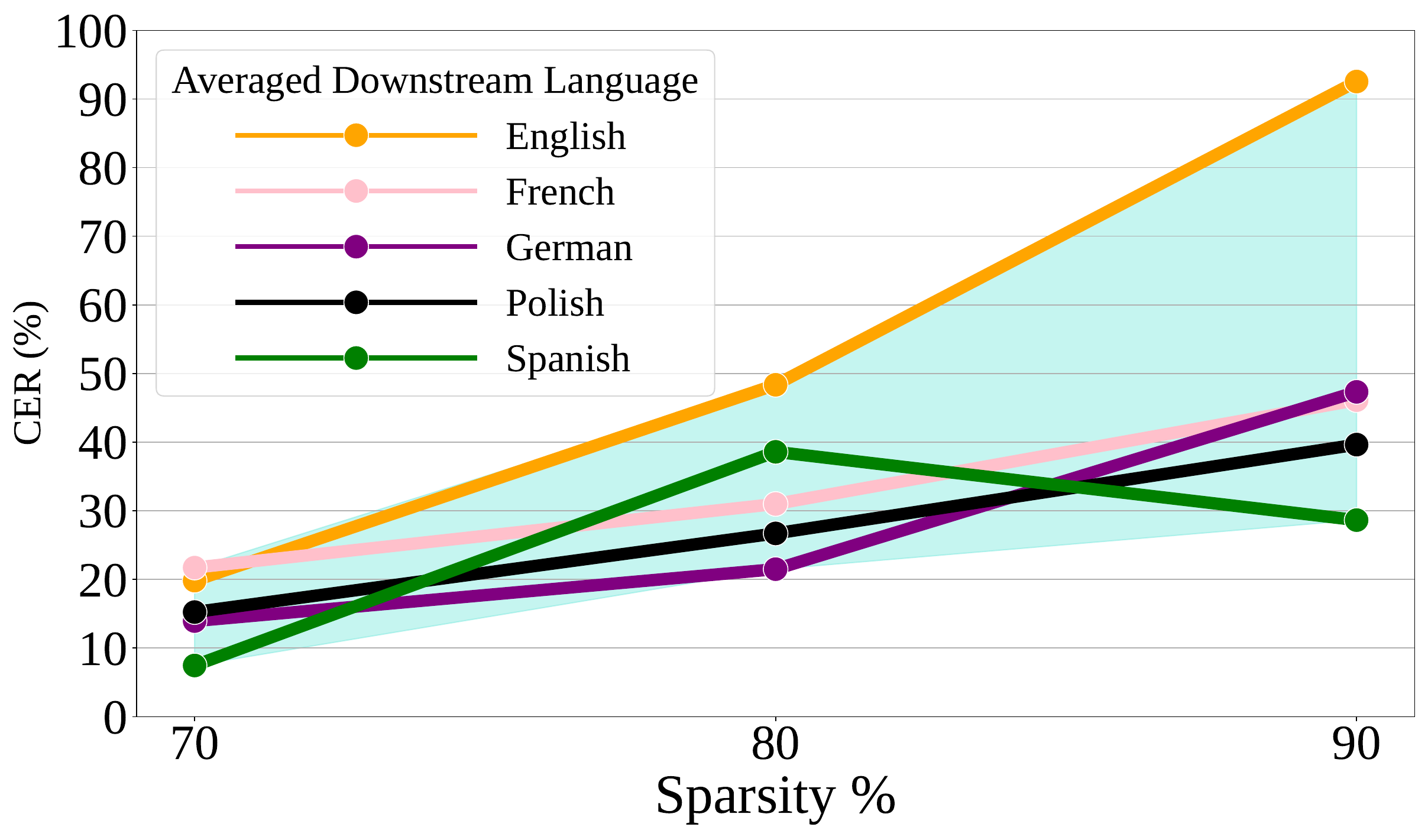}
  \caption{\textbf{Mean average results for each downstream language} each language is trained from an upstream model fine-tuned to each of the four other languages, these results are then averaged and plotted at 70\%, 80\% and 90\% sparsity. Results when the downstream and upstream languages match are not included in the averaging\\[-4ex]}
  \label{fig:AveragedLanguages}
  
\end{figure}

\newpage
Finally with Figure \ref{fig:AveragedLanguages} we measure the average CER for each language across all upstream models except for when the upstream and downstream languages match. For example, English refers in Figure \ref{fig:AveragedLanguages} to the mean average CER of the French, German, Polish and Spanish upstream models when fine-tuned to English. Figure \ref{fig:AveragedLanguages} shows English as having the highest CER at 80\% and 90\% sparsity and the second highest after French at 70\% sparsity. This is especially apparent at 90\% sparsity when the average CER for English is 92.55\% and the next lowest is German at 47.33\%. These results show English again as the outlier, here however it is the language that causes the highest number of errors among the other languages tested. 

\subsection{Evaluation of the English Subnetwork}
In Table \ref{tab:70perc90perc_English_Mask_Results} we compare the results of the English language-specific subnetwork to the Spanish language-specific subnetwork by fine-tuning to Asturian and Xhosa as well as English and Spanish. We compare the performance from the English subnetwork to the Spanish subnetwork at 70\% and 90\% sparsity. We also test how well Spanish and English fine-tuned models perform when pruned with subnetworks generated from each of the other languages. En / Es is the English fine-tuned model pruned to the Spanish subnetwork and Es / En is the Spanish fine-tuned model pruned to the English subnetwork. In both cases, the English upstream model pruned with the English subnetwork has the lowest average CER.

Finally, in Table \ref{tab:90perc_EnglishSpanish_Mixed_Mask} we show that combining all surviving weights from both English and Spanish subnetworks, noted in Table \ref{tab:90perc_EnglishSpanish_Mixed_Mask} as subnetwork En / Es, does not decrease CER more than the English single-language subnetwork alone. While the error on English is reduced on the Spanish upstream model, it is still 3\% higher in CER than the upstream English model with the English subnetwork.

\begin{table}[t]
\centering
\begin{tabular}{llll}
\cline{2-4} \\[-1.5ex] 
 & \multicolumn{3}{c}{\textbf{CER (\%) at 90\% Sparsity }} \\ \cline{2-4}{} \\[-1.5ex]
 & \multicolumn{3}{c}{\textbf{Language}} \\[0.5ex] \hline
\multicolumn{1}{l|}{\textbf{Model / Subnet}} & \multicolumn{1}{l}{\textbf{En}} & \multicolumn{1}{l}{} & \multicolumn{1}{l}{\textbf{Es}}  \\ \hline
\multicolumn{1}{r|}{\textbf{En / En}} & \multicolumn{1}{l}{\textbf{31.79}} & \multicolumn{1}{l}{} & \multicolumn{1}{l}{28.51}   \\ 
\multicolumn{1}{r|}{\textbf{Es / Es}} & \multicolumn{1}{l}{97.97} & \multicolumn{1}{l}{} & \multicolumn{1}{l}{\textbf{23.33}}  \\ 
\multicolumn{1}{r|}{\textbf{En / EnEs}} & \multicolumn{1}{l}{34.88} & \multicolumn{1}{l}{} &\multicolumn{1}{l}{27.74} \\ 
\multicolumn{1}{r|}{\textbf{Es / EnEs}} & \multicolumn{1}{l}{40.94} & \multicolumn{1}{l}{} & \multicolumn{1}{l}{23.86}   \\ \hline 
\end{tabular}
\caption{\textbf{Upstream English and Spanish models trained with mixed weight subnetworks at 90\% sparsity} we test the effect of combining the surviving weights across Spanish and English subnetworks after pruning at 90\% sparsity\\[-8ex]}
\label{tab:90perc_EnglishSpanish_Mixed_Mask}
\end{table}

\begin{figure}[b]
  \centering
  \includegraphics[width=\linewidth]{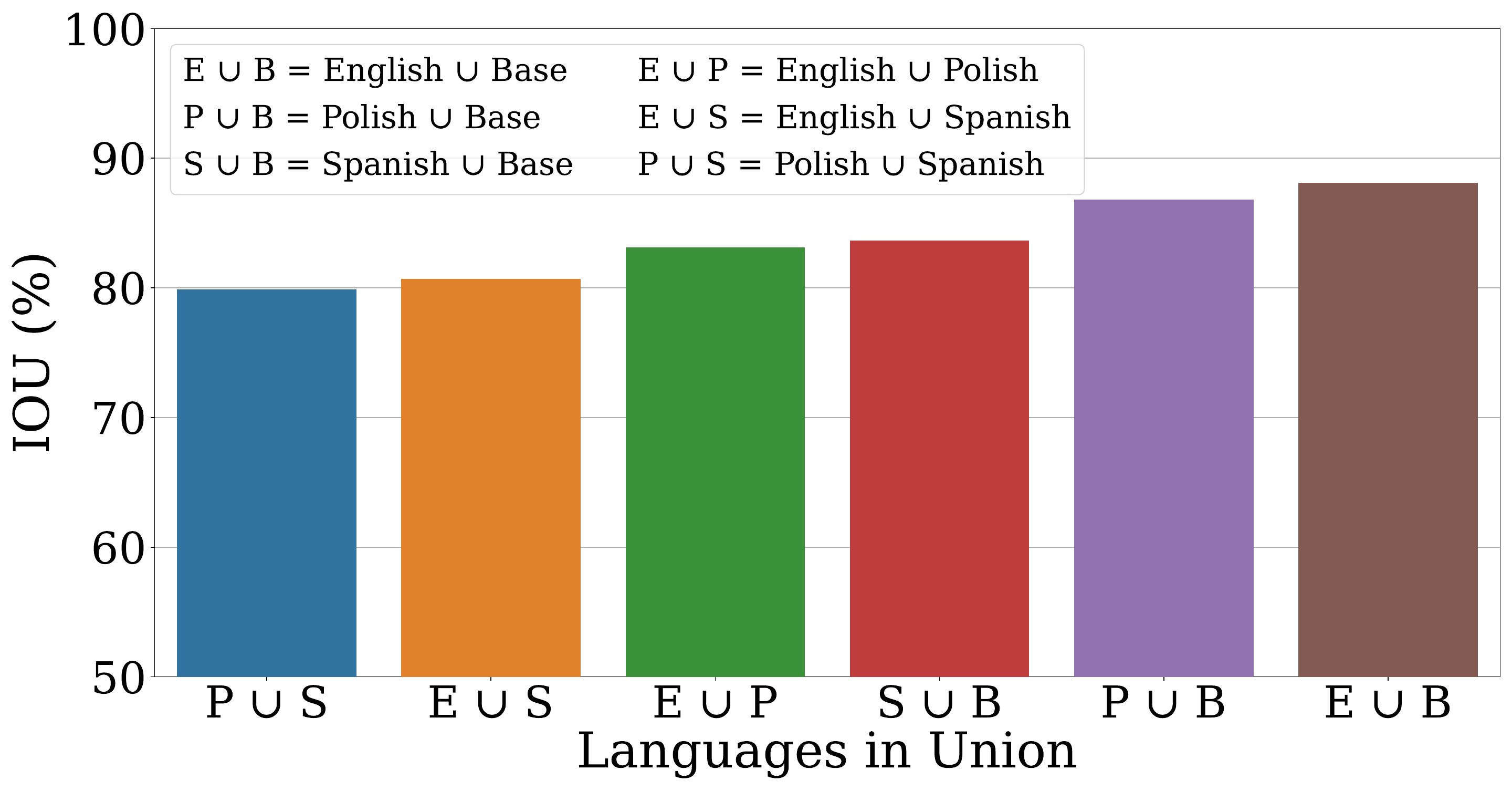}
  \caption{\textbf{Intersection Over Union (IOU) at 90\% sparsity} IOU overlap of weights in subnetworks between English, Spanish and Polish subnetworks and the subnetwork found in the base model before fine-tuning. All subnetworks pruned to 90\%.}
  \label{fig:IOU90perc}
\end{figure}

\begin{table*}
\centering
\begin{tabular}{lllllllllll}
& \\ \cline{2-11} \\[-2.3ex] 
& \multicolumn{5}{c}{\textbf{CER (\%) at 70\% Sparsity}} & \multicolumn{5}{|c}{\textbf{CER (\%) at 90\% Sparsity}} \\[0.75ex] \cline{2-11} \\[-2.35ex]
& \multicolumn{5}{c}{\textbf{Language}}  & \multicolumn{5}{|c}{\textbf{Language}} \\[0.5ex] \hline
\multicolumn{1}{l|}{\textbf{Model / Subnet}} & \multicolumn{1}{l}{\textbf{En}} & \multicolumn{1}{l}{\textbf{Es}} & \multicolumn{1}{l}{\textbf{As}} & \multicolumn{1}{l}{\textbf{Xh}} & \multicolumn{1}{l}{\textbf{Avg}} & \multicolumn{1}{|l}{\textbf{En}} & \multicolumn{1}{l}{\textbf{Es}} & \multicolumn{1}{l}{\textbf{As}} & \multicolumn{1}{l}{\textbf{Xh}} & \multicolumn{1}{l}{\textbf{Avg}} \\ \hline
\multicolumn{1}{r|}{\textbf{En / En}} & \multicolumn{1}{l}{\textbf{11.74}} & \multicolumn{1}{l}{6.61} & \multicolumn{1}{l}{\textbf{11.53}} & \multicolumn{1}{l}{\textbf{10.57}} & \textbf{10.11} & \multicolumn{1}{|l}{\textbf{31.79}} & \multicolumn{1}{l}{28.51} & \multicolumn{1}{l}{\textbf{33.31}} & \multicolumn{1}{l}{{23.06}} & \textbf{29.17} \\ 
\multicolumn{1}{r|}{\textbf{En / Es}} & \multicolumn{1}{l}{12.74} & \multicolumn{1}{l}{15.46} & \multicolumn{1}{l}{21.34} & \multicolumn{1}{l}{21.65} & 17.8 & \multicolumn{1}{|l}{44} & \multicolumn{1}{l}{28.86} & \multicolumn{1}{l}{34.61} & \multicolumn{1}{l}{{22.77}} & 32.56 \\ 
\multicolumn{1}{r|}{\textbf{Es / Es}} & \multicolumn{1}{l}{{28.28}} & \multicolumn{1}{l}{10.87} & \multicolumn{1}{l}{16.96} & \multicolumn{1}{l}{20.55} & 19.17 & \multicolumn{1}{|l}{{97.97}} & \multicolumn{1}{l}{\textbf{23.33}} & \multicolumn{1}{l}{33.49} & \multicolumn{1}{l}{\textbf{21.48}} & 44.07 \\ 
\multicolumn{1}{r|}{\textbf{Es / En}} & \multicolumn{1}{l}{14.11} & \multicolumn{1}{l}{\textbf{5.31}} & \multicolumn{1}{l}{27.4} & \multicolumn{1}{l}{23.25} & 17.52 & \multicolumn{1}{|l}{42.5} & \multicolumn{1}{l}{{26.52}} & \multicolumn{1}{l}{38.27} & \multicolumn{1}{l}{22.61} & 32.48 \\ \hline
\end{tabular}
\caption{\textbf{Upstream English and Spanish models trained with exchanged subnetworks at 70\% and 90\% sparsity} we test the effect of pruning with a Spanish subnetwork on an English upstream model and an English subnetwork on a Spanish upstream model across multiple languages at 70\% and 90\% sparsity\\[-3.6ex]}
\label{tab:70perc90perc_English_Mask_Results}
\end{table*}

\subsection{Intersection Over Union}
The results in this section so far have shown that fine-tuning on the English subnetwork achieves lower error at high sparsities than any other subnetwork. To explore further how this behaviour is occurring we can use the Intersection Over Union (IOU) equation from \cite{PARP}. IOU is a measure of how many weights overlap between two subnetworks. Figure \ref{fig:IOU90perc} shows the overlap of saved weights between two subnetworks when a model is pruned to 90\%. We test IOUs between English, Polish and Spanish subnetworks. Figure \ref{fig:IOU90perc} shows the lowest IOU at 80.67\% when comparing the English and Spanish subnetworks. This corroborates the findings in \cite{PARP} that discovered high overall overlap between different language-specific subnetworks. However, given our findings in Section \ref{sec:Results} we can surmise that the remaining weights not overlapping have significant influence over downstream training. We also see in Figure \ref{fig:IOU90perc} that the English language subnetwork has the most saved weights in common with the base XLS-R model. Adversely the Spanish subnetwork has the least in common with the base XLS-R weights. This may imply that fine-tuning XLS-R to English requires the least change in value for weights from the base model, prior to fine-tuning, when compared to other languages.

\begin{figure}[h]
  \centering
  \includegraphics[width=\linewidth]{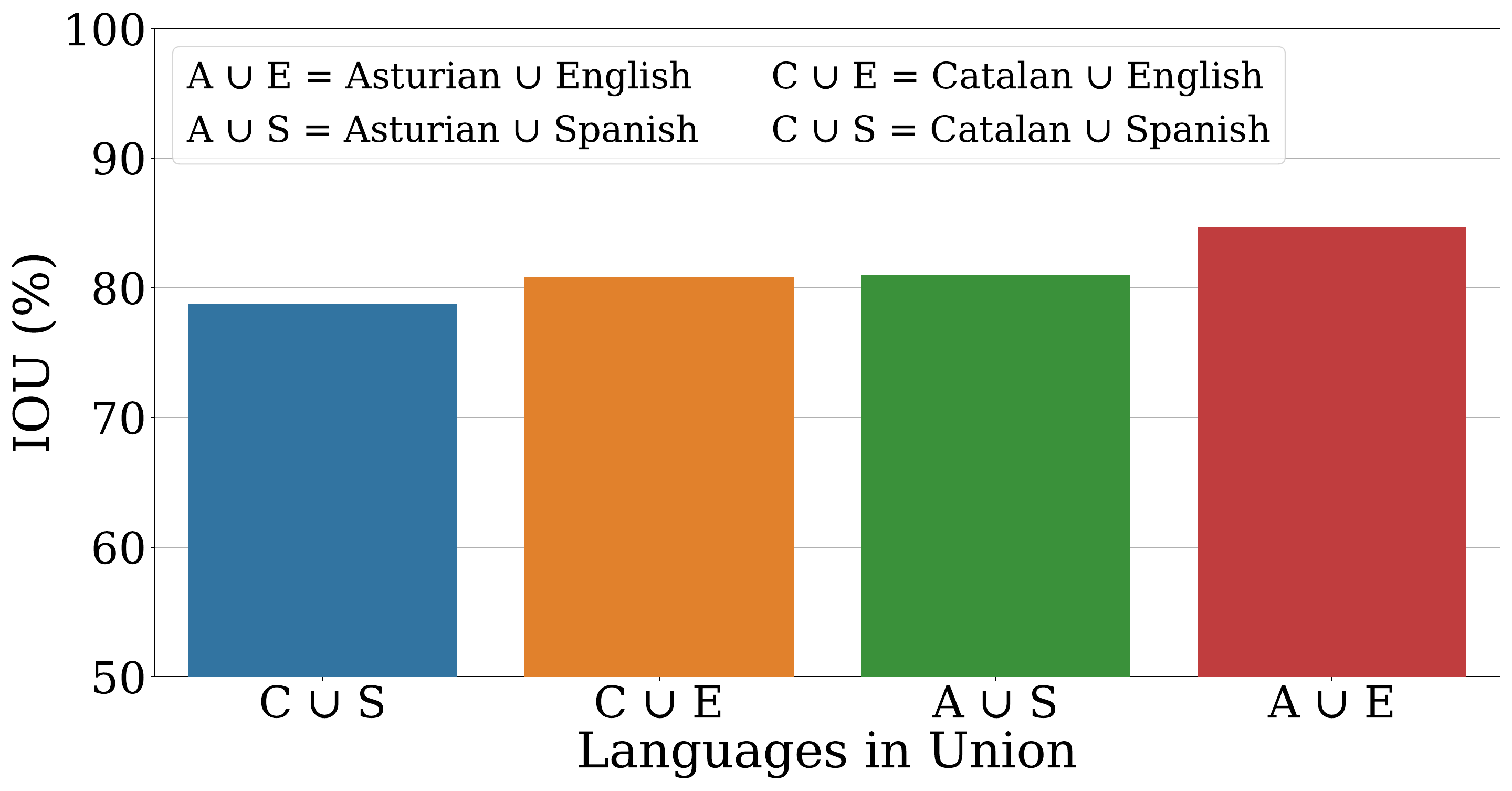}
  \caption{\textbf{Intersection Over Union (IOU) at 90\% sparsity for Spanish languages and English} Overlap of weights in subnetworks between English with Asturian and Catalan, then Spanish with Asturian and Catalan. Subnetworks pruned to 90\%.\\[-5ex]}
  \label{fig:IOU90perc_ACSvsE}
\end{figure}

Finally Figure \ref{fig:IOU90perc_ACSvsE} shows the IOUs calculated between four languages: English, Spanish, Catalan and Asturian. Spanish, Catalan and Asturian are all languages of Spain and all Latin languages whereas English is a Germanic language \cite{bauer2007linguistics}. Spanish and English are high-data languages in the XLS-R pretraining data, Catalan is a low-data language and Asturian is unseen in the XLS-R pretraining data. We see in Figure \ref{fig:IOU90perc_ACSvsE} that both Asturian and Catalan subnetworks have higher overlap in saved weights with the English language subnetwork than the Spanish subnetwork. Catalan has an IOU of 78.75\% with Spanish and 80.84\% with English and Asturian has an IOU of 81.01\% with Spanish and 84.66\% with English. This suggests that when learning low-data or new languages XLS-R is making use of the weights specific to English more than it is using weights specific to Spanish. This shows that XLS-R builds on weights learned for English when learning new or low-resource languages regardless of their language families.

\section{DISCUSSION}
Throughout Section \ref{sec:Results} we see the English subnetworks in XLS-R producing lower Character Error Rates (CER) when fine-tuned across all languages when compared to other language-specific subnetworks. Previous studies of language-specific subnetworks make the assumption that a subnetwork generated for a language will be the most efficient to further train on data from the same language or a linguistically related language \cite{Lu_Language_adap_xlsr, yang2023learning}. However, our experiments show that when pretraining data is imbalanced, this is not the case. The results in this paper instead show that regardless of linguistic content lower error will always be achieved by training on weights generated for the language with the highest data in pretraining, in this case English. The English dominance of XLS-R's performance is true for Germanic languages, such as English and German, but also languages from other Indo-European language families and one South African language. 

Through the use of the Intersection Over Union (IOU) we next analyse where overlap between the weights of different language-specific subnetworks occurs. We find the minimum overlap to be 78.75\%, which broadly corroborates previous results showing high overall overlap \cite{PARP}. However, given our previous findings, the remaining weights that are specific to one language do appear to impact downstream accuracy significantly. We compare each subnetwork with the base weights of XLS-R after pretraining but before fine-tuning. This shows us that the English subnetwork has the least deviation from the base model. This implies that the reliance on English has been taught to the model at the pretraining stage and not during downstream fine-tuning. Additionally, we also measure the overlap between subnetworks generated from two languages from Spain, Catalan and Asturian, to the English and Spanish subnetworks. We find that both Catalan and Asturian language-specific subnetworks have more overlap in saved weights with the English language-specific subnetwork. This reinforces our findings that when fine-tuning, XLS-R weights learned in pretraining from English language data are more impactful to training. This is regardless of the linguistic relation to English the downstream training data has.

This paper shows that Self-Supervised Learning (SSL) in Automatic Speech Recognition (ASR) can bias fine-tuning tasks towards weights learned from the data domains most present in its pretraining data. Open-source SSL ASR pretraining data is highly imbalanced towards a small number of languages. These models are over-reliant on the features learned from those languages which leads them to ignore linguistic relationships when fine-tuning to unseen or low-data languages. Previous research \cite{zhang2023fast} has shown that downstream tasks can benefit from linguistically guided pretraining so fine-tuning on linguistically unrelated features is both unintuitive and inefficient. Given this, we recommend balancing the pretraining data by language and linguistic relationships. When deploying pretrained open-source models researchers must carefully analyse the data the model has been pretrained with. Imbalanced pretraining data that is abundant in just one or a few languages, but limited across many others, is not best suited for efficient multilingual SSL ASR.

\section{ACKNOWLEDGEMENTS}
\textit{\footnotesize This work was conducted with the financial support of the Science Foundation Ireland Centre for Research Training in Digitally-Enhanced Reality (d-real) under Grant No. 18/CRT/6224 and ADAPT SFI Research Centre under Grant No. 13/RC/2106 P2. For the purpose of Open Access, the author has applied a CC BY public copyright licence to any Author Accepted Manuscript version arising from this submission} 

\newpage
\bibliographystyle{IEEEtran}
\bibliography{mybib}

\begin{thebibliography}{10}
\providecommand{\url}[1]{#1}
\csname url@samestyle\endcsname
\providecommand{\newblock}{\relax}
\providecommand{\bibinfo}[2]{#2}
\providecommand{\BIBentrySTDinterwordspacing}{\spaceskip=0pt\relax}
\providecommand{\BIBentryALTinterwordstretchfactor}{4}
\providecommand{\BIBentryALTinterwordspacing}{\spaceskip=\fontdimen2\font plus
\BIBentryALTinterwordstretchfactor\fontdimen3\font minus \fontdimen4\font\relax}
\providecommand{\BIBforeignlanguage}[2]{{%
\expandafter\ifx\csname l@#1\endcsname\relax
\typeout{** WARNING: IEEEtran.bst: No hyphenation pattern has been}%
\typeout{** loaded for the language `#1'. Using the pattern for}%
\typeout{** the default language instead.}%
\else
\language=\csname l@#1\endcsname
\fi
#2}}
\providecommand{\BIBdecl}{\relax}
\BIBdecl

\bibitem{babu2021xlsr}
A.~Babu, C.~Wang, A.~Tjandra, K.~Lakhotia, Q.~Xu, N.~Goyal, K.~Singh, P.~von Platen, Y.~Saraf, J.~Pino, A.~Baevski, A.~Conneau, and M.~Auli, ``{XLS-R}: Self-supervised cross-lingual speech representation learning at scale,'' 2021.

\bibitem{whisper}
A.~Radford, J.~W. Kim, T.~Xu, G.~Brockman, C.~McLeavey, and I.~Sutskever, ``Robust speech recognition via large-scale weak supervision,'' in \emph{International Conference on Machine Learning}.\hskip 1em plus 0.5em minus 0.4em\relax PMLR, 2023, pp. 28\,492--28\,518.

\bibitem{Huang_CrossLingual}
J.-T. Huang, J.~Li, D.~Yu, L.~Deng, and Y.~Gong, ``Cross-language knowledge transfer using multilingual deep neural network with shared hidden layers,'' in \emph{2013 IEEE International Conference on Acoustics, Speech and Signal Processing}, 2013, pp. 7304--7308.

\bibitem{Toshniwal_MultiLingLSTM}
S.~Toshniwal, T.~N. Sainath, R.~J. Weiss, B.~Li, P.~Moreno, E.~Weinstein, and K.~Rao, ``Multilingual speech recognition with a single end-to-end model,'' in \emph{2018 IEEE International Conference on Acoustics, Speech and Signal Processing (ICASSP)}, 2018, pp. 4904--4908.

\bibitem{wav2vec}
A.~Baevski, Y.~Zhou, A.~Mohamed, and M.~Auli, ``wav2vec 2.0: A framework for self-supervised learning of speech representations,'' \emph{Advances in neural information processing systems}, vol.~33, pp. 12\,449--12\,460, 2020.

\bibitem{hsu2021hubert}
W.-N. Hsu, B.~Bolte, Y.-H.~H. Tsai, K.~Lakhotia, R.~Salakhutdinov, and A.~Mohamed, ``Hubert: Self-supervised speech representation learning by masked prediction of hidden units,'' \emph{IEEE/ACM Transactions on Audio, Speech, and Language Processing}, vol.~29, pp. 3451--3460, 2021.

\bibitem{chen2022wavlm}
S.~Chen, C.~Wang, Z.~Chen, Y.~Wu, S.~Liu, Z.~Chen, J.~Li, N.~Kanda, T.~Yoshioka, X.~Xiao \emph{et~al.}, ``Wavlm: Large-scale self-supervised pre-training for full stack speech processing,'' \emph{IEEE Journal of Selected Topics in Signal Processing}, vol.~16, no.~6, pp. 1505--1518, 2022.

\bibitem{koenecke2020racial}
A.~Koenecke, A.~Nam, E.~Lake, J.~Nudell, M.~Quartey, Z.~Mengesha, C.~Toups, J.~R. Rickford, D.~Jurafsky, and S.~Goel, ``Racial disparities in automated speech recognition,'' \emph{Proceedings of the National Academy of Sciences}, vol. 117, no.~14, pp. 7684--7689, 2020.

\bibitem{towards_inclusive}
S.~Feng, B.~M. Halpern, O.~Kudina, and O.~Scharenborg, ``Towards inclusive automatic speech recognition,'' \emph{Computer Speech \& Language}, vol.~84, p. 101567, 2024.

\bibitem{sslbias_gender}
M.~Z. Boito, L.~Besacier, N.~Tomashenko, and Y.~Esteve, ``A study of gender impact in self-supervised models for speech-to-text systems,'' in \emph{Proc. Interspeech 2022}, 2022.

\bibitem{sslbias_speechrate}
Y.~Meng, Y.-H. Chou, A.~T. Liu, and H.-y. Lee, ``Don't speak too fast: The impact of data bias on self-supervised speech models,'' in \emph{ICASSP 2022 - 2022 IEEE International Conference on Acoustics, Speech and Signal Processing (ICASSP)}, 2022, pp. 3258--3262.

\bibitem{sslbias_Dutch}
M.~Fuckner, S.~Horsman, P.~Wiggers, and I.~Janssen, ``Uncovering bias in asr systems: Evaluating wav2vec2 and whisper for dutch speakers,'' in \emph{2023 International Conference on Speech Technology and Human-Computer Dialogue (SpeD)}, 2023, pp. 146--151.

\bibitem{zhang2023fast}
Z.~Zhang, W.~Wang, and Y.~Qian, ``Fast and efficient multilingual self-supervised pre-training for low-resource speech recognition,'' \emph{Proceedings of Interspeech. Dublin, Ireland}, 2023.

\bibitem{librispeech}
V.~Panayotov, G.~Chen, D.~Povey, and S.~Khudanpur, ``Librispeech: An asr corpus based on public domain audio books,'' in \emph{2015 IEEE International Conference on Acoustics, Speech and Signal Processing (ICASSP)}, 2015, pp. 5206--5210.

\bibitem{librilight}
J.~Kahn, M.~Rivière, W.~Zheng, E.~Kharitonov, Q.~Xu, P.~Mazaré, J.~Karadayi, V.~Liptchinsky, R.~Collobert, C.~Fuegen, T.~Likhomanenko, G.~Synnaeve, A.~Joulin, A.~Mohamed, and E.~Dupoux, ``Libri-light: A benchmark for asr with limited or no supervision,'' in \emph{ICASSP 2020 - 2020 IEEE International Conference on Acoustics, Speech and Signal Processing (ICASSP)}, 2020, pp. 7669--7673.

\bibitem{XLSR-53}
A.~Conneau, A.~Baevski, R.~Collobert, A.~Mohamed, and M.~Auli, ``{Unsupervised Cross-Lingual Representation Learning for Speech Recognition},'' in \emph{Proc. Interspeech 2021}, 2021, pp. 2426--2430.

\bibitem{MLS}
V.~Pratap, Q.~Xu, A.~Sriram, G.~Synnaeve, and R.~Collobert, ``{MLS: A Large-Scale Multilingual Dataset for Speech Research},'' in \emph{Proc. Interspeech 2020}, 2020, pp. 2757--2761.

\bibitem{distillation}
G.~Hinton, O.~Vinyals, and J.~Dean, ``Distilling the knowledge in a neural network,'' \emph{NIPS Deep Learning and Representation Learning Workshop}, 2015.

\bibitem{hu2021lora}
E.~J. Hu, Y.~Shen, P.~Wallis, Z.~Allen-Zhu, Y.~Li, S.~Wang, L.~Wang, and W.~Chen, ``Lora: Low-rank adaptation of large language models,'' 2021.

\bibitem{han2015learning}
S.~Han, J.~Pool, J.~Tran, and W.~Dally, ``Learning both weights and connections for efficient neural network,'' \emph{Advances in neural information processing systems}, vol.~28, 2015.

\bibitem{frankle2018LTH}
J.~Frankle and M.~Carbin, ``The lottery ticket hypothesis: Finding sparse, trainable neural networks,'' in \emph{International Conference on Learning Representations}, 2019.

\bibitem{PARP}
C.-I.~J. Lai, Y.~Zhang, A.~H. Liu, S.~Chang, Y.-L. Liao, Y.-S. Chuang, K.~Qian, S.~Khurana, D.~Cox, and J.~Glass, ``Parp: Prune, adjust and re-prune for self-supervised speech recognition,'' in \emph{Advances in Neural Information Processing Systems}, vol.~34, 2021, pp. 21\,256--21\,272.

\bibitem{Lu_Language_adap_xlsr}
Y.~Lu, M.~Huang, X.~Qu, P.~Wei, and Z.~Ma, ``Language adaptive cross-lingual speech representation learning with sparse sharing sub-networks,'' in \emph{ICASSP 2022 - 2022 IEEE International Conference on Acoustics, Speech and Signal Processing (ICASSP)}, 2022, pp. 6882--6886.

\bibitem{yang2023learning}
M.~Yang, A.~Tjandra, C.~Liu, D.~Zhang, D.~Le, and O.~Kalinli, ``Learning asr pathways: A sparse multilingual asr model,'' in \emph{ICASSP 2023-2023 IEEE International Conference on Acoustics, Speech and Signal Processing (ICASSP)}.\hskip 1em plus 0.5em minus 0.4em\relax IEEE, 2023, pp. 1--5.

\bibitem{wang-etal-2021-voxpopuli}
C.~Wang, M.~Riviere, A.~Lee, A.~Wu, C.~Talnikar, D.~Haziza, M.~Williamson, J.~Pino, and E.~Dupoux, ``{V}ox{P}opuli: A large-scale multilingual speech corpus for representation learning, semi-supervised learning and interpretation,'' in \emph{Proceedings of the 59th Annual Meeting of the Association for Computational Linguistics and the 11th International Joint Conference on Natural Language Processing (Volume 1: Long Papers)}, C.~Zong, F.~Xia, W.~Li, and R.~Navigli, Eds.\hskip 1em plus 0.5em minus 0.4em\relax Association for Computational Linguistics, Aug. 2021, pp. 993--1003.

\bibitem{FLEURS}
A.~Conneau, M.~Ma, S.~Khanuja, Y.~Zhang, V.~Axelrod, S.~Dalmia, J.~Riesa, C.~Rivera, and A.~Bapna, ``Fleurs: Few-shot learning evaluation of universal representations of speech,'' in \emph{2022 IEEE Spoken Language Technology Workshop (SLT)}, 2023, pp. 798--805.

\bibitem{bauer2007linguistics}
L.~Bauer, \emph{Linguistics Student's Handbook}.\hskip 1em plus 0.5em minus 0.4em\relax Edinburgh University Press, 2007.

\bibitem{Asturian_Muñiz-Cachón_2018}
C.~Muñiz-Cachón, ``Asturian,'' \emph{Journal of the International Phonetic Association}, vol.~48, no.~2, p. 231–241, 2018.

\bibitem{FLEURS_XLSR_WHISPER}
A.~Rouditchenko, S.~Khurana, S.~Thomas, R.~Feris, L.~Karlinsky, H.~Kuehne, D.~Harwath, B.~Kingsbury, and J.~Glass, ``{Comparison of Multilingual Self-Supervised and Weakly-Supervised Speech Pre-Training for Adaptation to Unseen Languages},'' in \emph{Proc. INTERSPEECH 2023}, 2023, pp. 2268--2272.

\bibitem{ding2022audio}
S.~Ding, T.~Chen, and Z.~Wang, ``Audio lottery: Speech recognition made ultra-lightweight, noise-robust, and transferable,'' in \emph{International Conference on Learning Representations}, 2022.

\bibitem{Losses_NEURIPS2022_83d349b6}
Y.~Fu, Y.~Zhang, K.~Qian, Z.~Ye, Z.~Yu, C.-I.~J. Lai, and C.~Lin, ``Losses can be blessings: Routing self-supervised speech representations towards efficient multilingual and multitask speech processing,'' in \emph{Advances in Neural Information Processing Systems}, S.~Koyejo, S.~Mohamed, A.~Agarwal, D.~Belgrave, K.~Cho, and A.~Oh, Eds., vol.~35.\hskip 1em plus 0.5em minus 0.4em\relax Curran Associates, Inc., 2022, pp. 20\,902--20\,920.

\bibitem{Unstructured_Speech}
Y.~Peng, K.~Kim, F.~Wu, P.~Sridhar, and S.~Watanabe, ``Structured pruning of self-supervised pre-trained models for speech recognition and understanding,'' in \emph{ICASSP 2023 - 2023 IEEE International Conference on Acoustics, Speech and Signal Processing (ICASSP)}, 2023, pp. 1--5.

\end{thebibliography}
\end{document}